\documentclass[journal]{IEEEtran}
\ifCLASSINFOpdf
\usepackage[pdftex]{graphicx}
\graphicspath{{bio_pictures/}{figures/}}
\else
\fi
\usepackage{amsmath}
\usepackage{amssymb}
\usepackage{algorithmic}
\usepackage[ruled,noline]{algorithm2e} 

\makeatletter
\newcommand{\removelatexerror}{\let\@latex@error\@gobble}
\makeatother

\usepackage{multirow}
\usepackage{listings}

\begin{document}
%
\title{Diagonal Memory Optimisation for Machine Learning on Micro-controllers}
%
%
%

\author{P.~Blacker,~\IEEEmembership{}
        C.~P.~Bridges,~\IEEEmembership{}
        S.~Hadfield~\IEEEmembership{}
\thanks{P. Blacker and C. P. Bridges are with the Surrey Space Centre, Surrey University, UK.}
\thanks{S. Hadfield is with CVSSP, University of Surrey, UK.}}

%
%

\markboth{ }%
{Shell \MakeLowercase{\textit{et al.}}: Diagonal Memory Optimisation for Machine Learning on Micro-controllers}
%



\maketitle

\begin{abstract}
As machine learning spreads into more and more application areas, micro controllers and low power CPUs are increasingly being used to perform inference with machine learning models. The capability to deploy onto these limited hardware targets is enabling machine learning models to be used across a diverse range of new domains. Optimising the inference process on these targets poses different challenges from either desktop CPU or GPU implementations, where the small amounts of RAM available on these targets sets limits on size of models which can be executed. Analysis of the memory use patterns of eleven machine learning models was performed. Memory load and store patterns were observed using a modified version of the Valgrind debugging tool, identifying memory areas holding values necessary for the calculation as inference progressed. These analyses identified opportunities optimise the memory use of these models by overlapping the input and output buffers of individual tensor operations. Three methods are presented which can calculate the safe overlap of input and output buffers for tensor operations. Ranging from a computationally expensive approach with the ability to operate on compiled layer operations, to a versatile analytical solution which requires access to the original source code of the layer. The diagonal memory optimisation technique is described and shown to achieve memory savings of up to 34.5\% when applied to eleven common models. Micro-controller targets are identified where it is only possible to deploy some models if diagonal memory optimisation is used.
\end{abstract}

%
\IEEEpeerreviewmaketitle

\section{Introduction}\label{sec:introduction}

\begin{figure}[!t]
\centering
\includegraphics[width=2.5in]{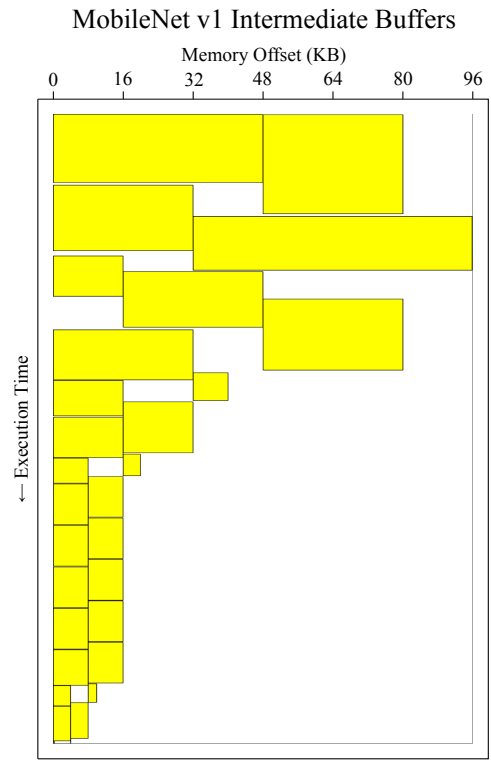}
\caption{Intermediate tensor buffer locations in for the example model (MobileNet v1 0.25 128 quantised).Location within the tensor area memory is shown on the x-axis while the scope of each buffer from first to last use is shown on the y-axis.}
\label{fig:mobile_net_blocks}
\end{figure}

\begin{figure*}[!tb]
\centering
\includegraphics[width=6in]{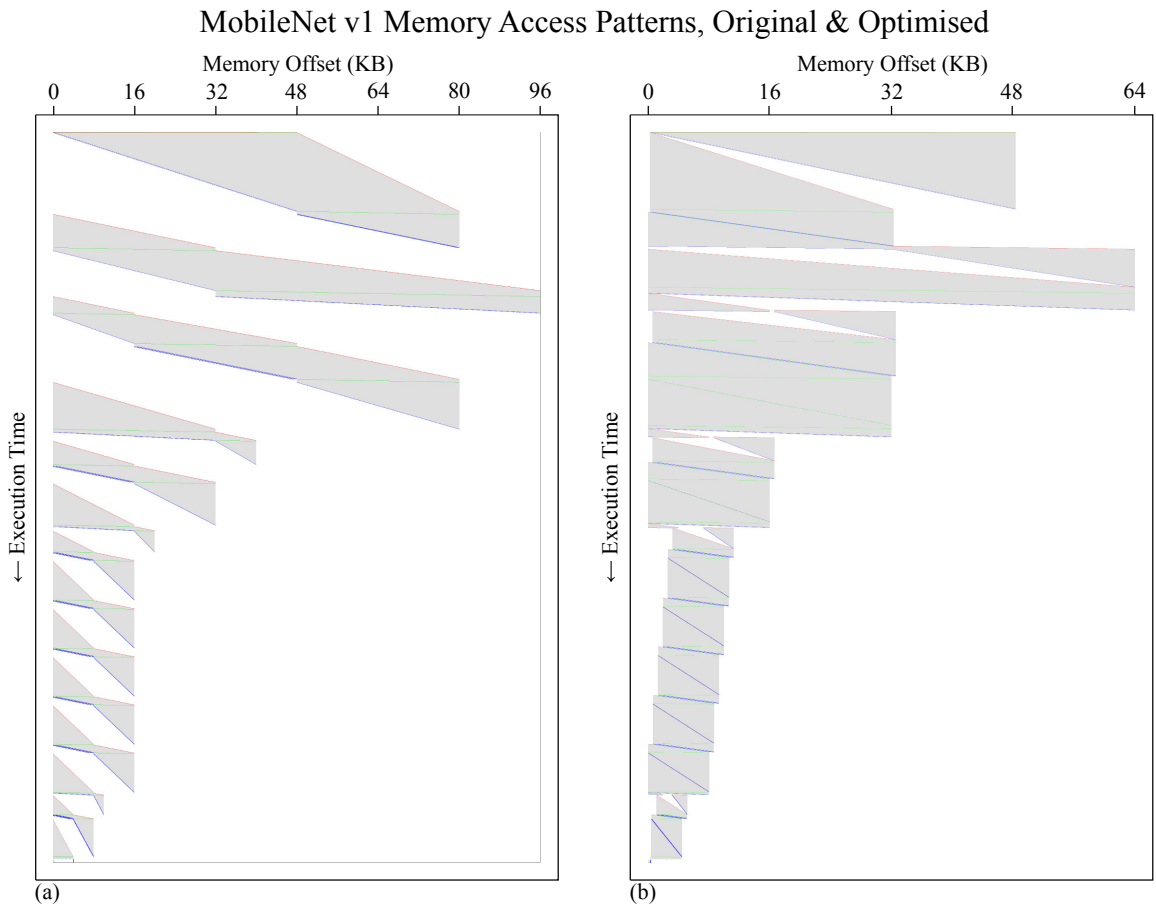}
\caption{Intermediate buffer memory access pattern for the example model (MobileNet v1 0.25 128 quantised). In use areas shown in grey, load, store, and update events in red, blue, and green respectively. Plot a shows the memory access pattern when the original heap allocation strategy is used to allocate intermediate buffers, large areas of unused memory can be seen indicating that a more optimal use of memory is possible. Plot b shows the memory access pattern of the same model with intermediate buffers allocated using diagonal memory optimisation, in-use memory is packed more densely allowing the size of the tensor area to be reduced.}
\label{fig:mobile_net_traces}
\end{figure*}

\IEEEPARstart{L}{imited} processing power available on embedded CPUs is a hindrance when running machine learning (ML) models, but if longer execution time is acceptable then it is still possible to perform inference. The amount of RAM however places a hard limit on the size of models which can be executed. If intermediate tensor values do not fit in the available memory on a target, there is no straightforward method to be able to execute the model. Taking an example network which is already optimised to be small, MobileNet v1.0 0.25 128, quantised to 8 bits \cite{howard2017mobilenets}. When inference is being performed with this model the second 2D convolution operation needs 32 KB input and 64 KB output buffers. This operation sets the peak RAM requirement for inference with this model at 96 KB as can be seen in the intermediate buffer allocation pattern shown in Figure \ref{fig:mobile_net_blocks}. This figure displays the location of intermediate tensor buffers, where location within the tensor arena is shown on the x-axis while its scope (first use to final use) is shown on the y-axis. Tensorflow Lite Micro \cite{tinyml} uses a monolithic memory region known as the tensor arena to hold intermediate tensor values. If no buffer pre-allocation information is provided alongside the model then a heap allocation strategy is used by default.

To investigate opportunities for reducing the peak memory usage and therefore tensor area size needed by ML models a customised version of the Valgrind debugging tool \cite{nethercote2007valgrind} was developed, this tool can observe memory read and write operations within the tensor arena as the model is being executed. This analysis determines the areas of memory which are holding values which are later read and used for calculations, and redundant memory where the value is never read. Figure \ref{fig:mobile_net_traces} a shows a memory trace produced by this tool of the same MobileNet implementation described in Figure \ref{fig:mobile_net_blocks}. Exposing the internal workings of each operation in this way an opportunity can be clearly seen to reduce the peak memory requirement by overlapping the input and output buffers of operations.

This approach of diagonal memory optimisation can significantly reduce the peak memory demands of machine learning models but is only possible if the underlying layer implementations are known and functions exist to determine the safe overlap offset of a specific layer instance. This requires an architectural change from pre-allocation schemes currently used by Tensorflow Lite Micro and uTensor from ARM \cite{utensor}, where intermediate buffers are allocated without any knowledge of the layer implementations which will eventually use them.

Calculating the safe input/output buffer overlap for tensor operations is not a task that has been described in literature before. Yet it is an essential capability of tools that can automatically leverage this optimisation strategy. This work investigates three approaches to the calculation of the buffer safe overlap ($O_s$) and presents a set of analytical solutions for common tensor operations. These solutions are evaluated by generating buffer pre-allocation patterns for eleven notable ML models, memory reductions of up-to 33\% were achieved. Figure \ref{fig:mobile_net_traces} b illustrates how diagonal memory optimisation reduces the amount of memory MobileNet requires by packing in-use areas of memory together more densely than block level optimisers.

Several existing memory optimisation strategies for edge ML implementations are described alongside diagonal memory optimisation in section \ref{sec:mem_opt}. It is shown that diagonal memory optimisation is complimentary most existing approaches enabling even greater savings. The task of computing $O_s$ for any layer implementation is investigated in Section \ref{sec:calc_offsets} and three different methods are presented. Diagonal memory optimisation with performance optimised layer implementations is the discussed with reference to vectorisation and multi-threading.

The optimised buffer allocation patterns generated in this work have been verified as safe using an open-source tool TFMin \cite{tf_min} released by the authors. This tool generates ANSI c implementations of ML models from Tensorflow sessions and TFLite flatbuffers, which utilise fixed pre-allocated memory patterns. Example code is provided which performs the Diagonal Memory Optimisation described in this paper on the MobileNet v1 and v2 models \cite{howard2017mobilenets}, and verifies the correct estimates are produced.

\section{Optimising Memory use of Edge ML}\label{sec:mem_opt}
Machine learning models are in essence graph functions made up of large tensor operations. These operations are executed sequentially to perform inference and sufficient memory must be available to store the intermediate values needed during this process. Figure 1 shows the intermediate buffer locations and scopes for a MobileNet, allocated using a heap approach. In this case the peak memory requirement is defined by the third and four buffers which are both needed at the same time taking 96 KB combined. Looking at these buffer allocations it does not immediately seem possible to reduce this memory requirement further, however several methods do exist to achieve this. Complimentary to those we introduce diagonal memory optimisation which can further reduce peak memory requirements.

\subsection{Operation Splitting}\label{sec:operation_splitting}
Sets of operations requiring large intermediate buffers can in some instances be split up into smaller operations and executed sequentially. This has the effect that fewer intermediate values need to be stored at the same time, at the cost of some tensor elements being calculated more than once. This method can be demonstrated using MobileNet v1 where the second and third operations (a 2D convolution \& depthwise convolution) between them process a 32 KB tensor into a 16 KB tensor. However the intermediate tensor between these two operations takes 64 KB increasing the peak memory requirement of this model.

Due to the small kernel sizes used in this model the receptive field of each element in the final 16 KB tensor is a $3 \times 3 \times depth$ patch of the 32 KB input tensor. If these two operations are split into four pairs of operations where each pair computes a quarter of the final output tensor then four consecutive intermediate tensors are needed of at most 18 KB. This reduces the peak memory requirement of the model to 66 KB, however due to the spatial overlap of the smaller intermediate tensors 6144 elements need to be computed twice. The longer scope of the input and output tensors means that this approach can not be combined with diagonal memory optimisation.

The authors have used this approach manually but have not found any formal descriptions in literature. Operation splitting has the useful ability to trade memory for processing time to varying degrees. There is scope for further work in this area to find methods to automatically analyse tensor graphs for potential savings and their associated performance penalties.

\subsection{Graph Serialisation}\label{sec:graph_serialisation}
The MobileNet example shown has a perfectly sequential tensor graph, however other networks such as Nasnet or DenseNet have more connected topologies. Operations in these connected graphs can be serialised into different valid orders which in some cases affects the number of intermediate tensors needed therefore impacting the peak memory requirement. The problem of serialising these graphs in such a way to minimise their memory requirement is NP-hard, so heuristic methods such as the BMS scheduler proposed by Sbîrlea et. al. \cite{sbirlea2014bounded} are needed.

This re-ordering of operations has been investigated by ARM in their micro ML tool uTensor \cite{utensor} and Blacker et. al in TFMin \cite{blacker2019rapid}. Only more complex connected ML models can be optimised using operation re-ordering meaning it can not be used with the smaller models often used on micro-controllers, however it can be used alongside diagonal memory optimisation.

\subsection{Operation Removal}\label{sec:operation_removal}
Element re-arrangement operations such as concatenate and pack are common in ML models and define the peak memory requirements of some ML models such as Squeezenet \cite{iandola2016squeezenet}. In most cases it is theoretically possible to remove these operations if the upstream operations are able to write their output directly into the aggregated tensor. This removes the need to store two identical, albeit different shaped copies of the same elements in memory.

The current architecture of Tensorflow Lite Micro is not capable of mapping tensor elements into memory in a way that enables this approach. However it could be added with a small change to the memory offset function. This approach can be combined with diagonal memory optimisation, although the change to the offset function does alter the computation of $O_s$ described in Section \ref{sec:calc_offsets}.

\subsection{Diagonal Memory Optimisation}\label{sec:diag_optimisation}
The diagonal memory optimisation technique proposed here works at a lower level than these graph based approaches and is complimentary to the graph serialization and operation removal techniques described. The observation that the input and output buffers of many tensor operation can be safely overlapped is utilised in cases where the input to an operation is not needed by later operations. Therefore the input buffer can be overwritten during the computation of the operation's output.
A heap based allocation approach was used to place intermediate tensor into the tensor arena in reverse execution order. This ordering was used because diagonal memory optimisation allows the start of the input buffer to overlap with the end of the output buffer. The result of the diagonal memory optimisation approach with buffer overlapping can be seen in Figure \ref{fig:mobile_net_traces} b.

Since buffers are allocated in reverse order this approach can only be used as a pre-allocation method before the inference process has started. Our results presented in Section \ref{sec:results} were generated using this reverse heap allocation approach with the analtyical method of computing the safe buffer overlap $O_s$ described in Section \ref{sec:analytical}.

\section{Calculating the Safe Buffer Overlap}\label{sec:calc_offsets}
To understand the definition of the safe buffer overlap three methods to determine $O_s$ are presented. Our initial work debugging compiled networks is described first, followed by a more efficient algorithmic approach. Ultimately a method to derive analytical lower bounds of $O_s$ is presented along with a discussion of their precision. Finally we discuss the effect of performance optimisation techniques on the utility of buffer overlapping.

\subsection{Definition of the Safe Buffer Overlap}\label{sec:overlap_definition}

\begin{figure}[!t]
\centering
\includegraphics[width=2.5in]{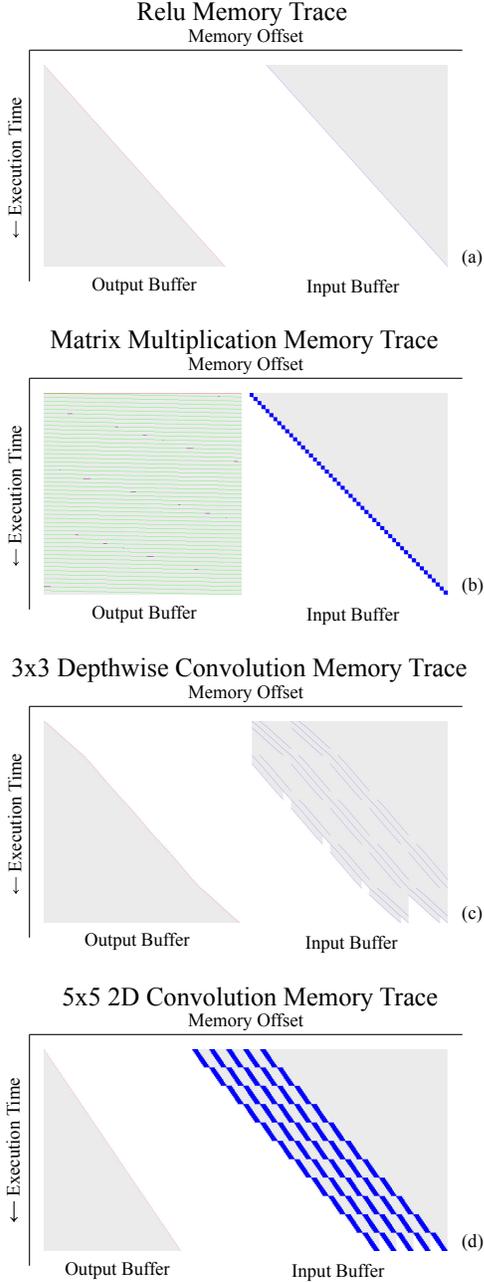}
\caption{Memory traces of four common ML tensor operations. (a) Rectified Linear Unit, (b) Matrix Multiplication, (c) Depthwise Convolution, (d) 2D Convolution. These traces only show the intermediate input \& output tensor buffers, omitting the filter and weight buffers.}
\label{fig:four_traces}
\end{figure}

It can be intuitively seen in Figure 3 that the input and output buffers of three of the four tensor operations can be overlapped a certain amount without any values in memory being clobbered. The type of operation, algorithm used to compute the result, size inputs, and parameters all determine its pattern of memory access and therefore the exact size of this safe overlap. 

Element-wise unary and binary operations such as the Relu shown in Figure \ref{fig:four_traces} a, have perfectly diagonal input and output patterns representing the ideal case where $O_s$ is simply the size of the output buffer. It is interesting to observe here, that in-place buffer re-use is actually a special case of diagonal memory optimisation. The matrix multiplication operation shown in Figure \ref{fig:four_traces} b represents the other extreme, the whole range of its output buffer is repeatedly updated until the final slice is processed. In this case the input and output buffers can not be overlapped at all. Depth-wise convolution and 2D convolution operations shown in Figures \ref{fig:four_traces} c \& d fall somewhere between these two extremes. 

\begin{figure}[!t]
\centering
\includegraphics[width=2.5in]{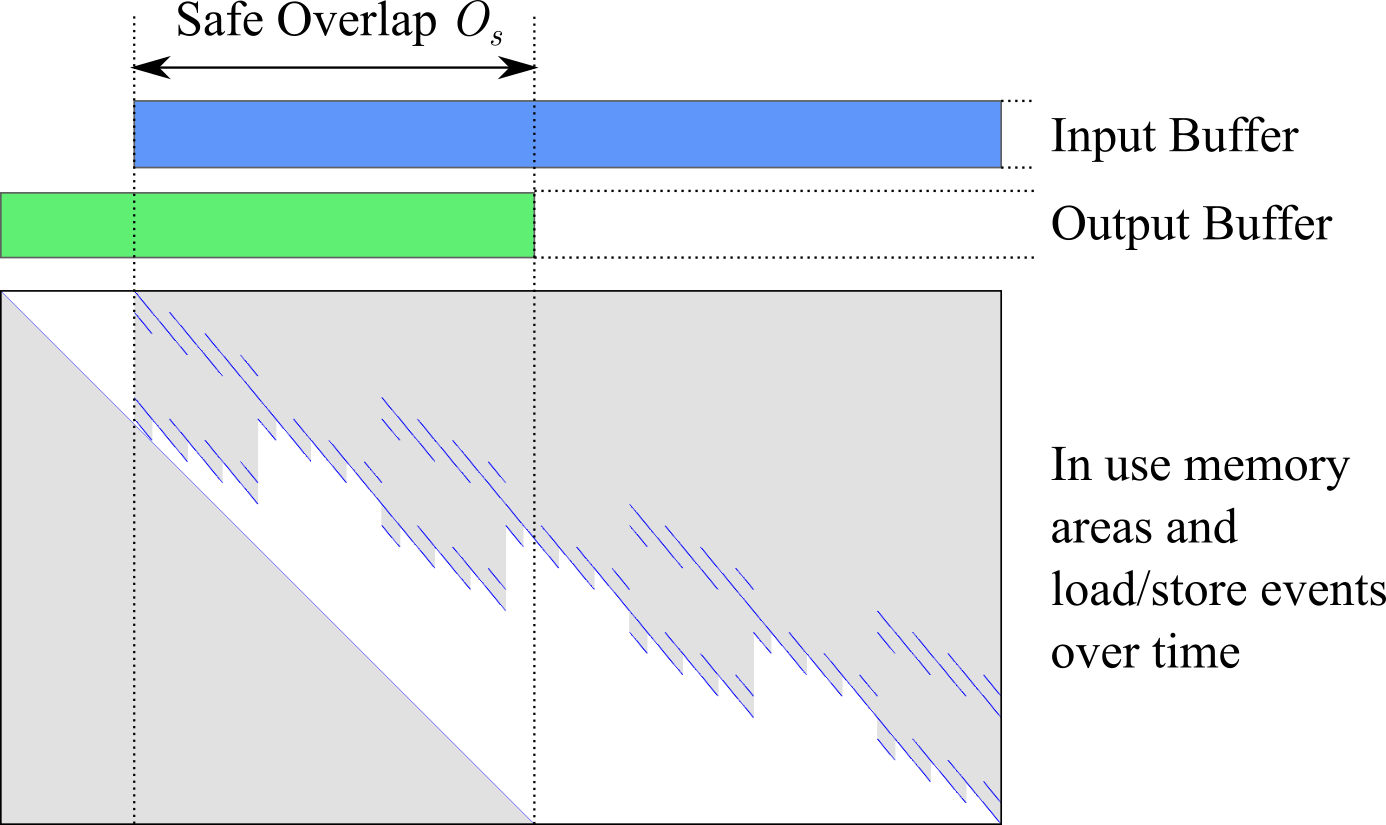}
\caption{Definition of the safe buffer overlap ($O_s$) metric, defined as the maximum overlap where no in-use areas of memory are clobbered.}
\label{fig:overlap_def}
\end{figure}

By convention implementations progress from lower indices to higher indices, for simplicity this work assumes that algorithms will always be processed in this direction. Although it is theoretically possible to make use of algorithms which can process in either direction, this work has not investigated this option. Safe overlap $O_s$ is formally defined as the maximum number of bytes that the start of the input buffer can be overlapped with the end of the output buffer without clobbering any values in memory, as shown in Figure \ref{fig:overlap_def}. The memory saved for each operation is identical to the buffer overlap $O_s$ itself. The process of determining $O_s$ and using the safe overlap when allocating intermediate buffers is the central new concept of diagonal memory optimisation.

\subsection{Bottom up Method}\label{sec:bottom_up}
Bottom up approaches such as the Valgrind \cite{nethercote2007valgrind} method described in Section \ref{sec:introduction} observe the load and store operations of a compiled operation as it is being debugged. The difference between this and a conventional memory trace is a mechanism to isolate the memory operations of the layer implementation from the rest of the compiled binary. The authors used a FIFO between the binary on test and the debugger itself to inform the debugger of the input and output buffer locations. A dedicated memory region within the binary on test was used to signal the start and end of each layers computation. This tool was used to record the memory access patterns of single layer operations and whole ML models, identifying the original opportunity for memory optimisation of MobileNet shown in Figure \ref{fig:mobile_net_traces} a.

The raw output of this introspection method is a set of memory events at 2D locations in time and buffer-offset, measured in instructions and bytes respectively. This type of raw output was used to produce the plots shown in Figure \ref{fig:four_traces} showing the memory access patterns of four types of operation. These events can also be processed to find the maximum safe overlap $O_s$ between the input and output buffers of the operation being analysed.

The advantage of the bottom-up method is that the layer implementation can be a black-box, even implementations in compiled libraries can be analysed meaning this approach can be used with many existing operations. There is however a requirement that memory read/write behaviour must be deterministic, which excludes multi-threaded implementations due to the non-deterministic nature of thread synchronisation. This limitation applies to diagonal memory optimisation itself which is discussed further in Section \ref{sec:effects_of_opt}. The disadvantage of this method is its high computational cost and the complexity of the process. Building dedicated test binaries and debugging layer implementations is a complex approach, the algorithmic and analytic methods described offer faster and more portable approaches to find $O_s$ but require access to original source code.

\subsection{Algorithmic Method}\label{sec:algorithmic}
The algorithmic method requires the development of a new algorithm to compute $O_s$ based upon the original layer implementation. This new algorithm removes the calculation of tensor values leaving only the calculation of buffer offsets where these values would have been read from and written to. 

The original implementation is analysed so that the number of write or update operations ($Steps$) on the output buffer can be determined. A new algorithm is then written which produces two arrays $minR$ \& $maxW$ each $Steps$ long. Where each element of $minR$ contains the minimum read offset of that step and all future steps, while each element of $maxW$ contains the maximum write offset of this step and all previous steps. $O_s$ can then be calculated using $O_s = O_b + minD$ where $O_b$ is the output buffer size and $minD$ is the minimum of $minR-maxW$ across the arrays. This core conceptual structure of the algorithmic method and is enough for a developer to write a new algorithm to compute for $O_s$ for any existing deterministic algorithm.

A practical demonstration of this method is presented using pseudo code of the depthwise 2D convolution reference implementation from Tensorflow lite shown in Algorithm \ref{alg:dw_conv}. Bias and activation functions have been omitted for clarity since they have no effect on the computation of $O_s$.

\begin{algorithm}\label{alg:dw_conv}
\SetAlgoLined
\caption{Depthwise 2D Convolution - Pseudo Code}
\For{$b = 0$ to $batches$}{
	\For{$outY = 0$ to $outputH$}{
		\For{$outX = 0$ to $outputW$}{
			\For{$ic = 0$ to $inputD$}{
				\For{$m = 0$ to $filterC$}{
					$total \gets 0$
					\For{$filterY = 0$ to $filterH$}{
						\For{$filterX = 0$ to $filterW$}{
							\If{input element in input tensor}{
								$F_o \gets$ [calc filter offset]
								$I_o \gets$ [calc input offset]
								$total \gets total + (filter[F_o] \times input[I_o])$
							}
						}
					}
					$O_o \gets$ [calc output offset]
					$output[O_o] \gets total$
				}			
			}
		}
	}
}
\end{algorithm}

In this case a single output element is computed within each iteration of the $5^{th}$ nested loop, therefore the $minR$ \& $maxW$ arrays need to be $batches\times outputH\times outputW\times inputD\times filterC$ elements long. The value of $minR$ can be found for each iteration as the minimum of all values computed within the $filterY$ \& $filterX$ loops, as long as a final reverse pass is performed to enforce the 'minimum of all future iterations' requirement. The value of $maxW$ for an iteration is the highest value of $O_o$ computed so far through the loops. These modifications are shown in Algorithm \ref{alg:os_computation}.

\begin{algorithm}\label{alg:os_computation}
\SetAlgoLined
\caption{Computation of $O_s$ - Pseudo Code}
$Steps\gets batches\cdot outputH\cdot outputW\cdot inputD\cdot filterC$
$minR = array(iCount)$
$maxW = array(iCount)$
$maxF_o = 0$
$it = 0$
\For{$b = 0$ to $batches$}{
	\For{$outY = 0$ to $outputH$}{
		\For{$outX = 0$ to $outputW$}{
			\For{$ic = 0$ to $inputD$}{
				\For{$m = 0$ to $filterC$}{
					$minR_o\gets +inf$
					\For{$filterY = 0$ to $filterH$}{
						\For{$filterX = 0$ to $filterW$}{
							\If{input element in input tensor}{
								$I_o \gets$ [calc input offset]
								$minR_o\gets min(minR_o, I_o)$
							}
						}
					}
					$minR[it]\gets minR_o$
					$O_o\gets$ [calc output offset]
					$maxW[it]\gets max(maxF_o, O_o)$
					$it\gets it + 1$
				}			
			}
		}
	}
}
$minR_o\gets +inf$
$minD\gets 0$
\For{$i = Steps$ to $0$}{
	$minR[i] = min(minR[i], minR_o$)
	$minD\gets min((minR[i] - maxW[i]), minD)$
}
$O_s = outputBufSize + minD$
\end{algorithm}

An implementation of the pseudo code above can be used to calculate the value of $O_s$ for any instance of the reference depthwise 2D convolution operation directly without the need to inspect the behaviour of a compiled layer. The pattern of code changes in the demonstration above can be applied to any single-threaded tensor operation converting it into an algorithm for the direct computation of $O_s$. In this specific example further inspection of the source code reveals that the values of $minR_o$ and $O_o$ calculated by the first set of loops will always be monotonic with respect to $it$. Therefore in this case the code could be simplified to a single set nested of loops. 

The algorithmic method is significantly faster and more convenient than the bottom-up method, however it still requires almost the full layer operation to be executed. Since $O_s$ is generated by complex algorithms it is difficult to generalise the solutions between different types of layer implementation. These shortcomings are addressed using by analytical method described in the following section.

\subsection{Analyical Method}\label{sec:analytical}
The Analytic approach derives an equation for a specific layer implementation which directly calculates \begin{math}O_s\end{math} for any instance of that layer. This approach requires the least computation time and more importantly is the least error-prone when translating between programming languages. We describe the approach taken to derive these analytical solutions for several common ML layer implementations. These analytical $O_s$ equations were then used to generate optimised buffer pre-allocations of the eleven test models shown in Section \ref{sec:results}.

It is important to note that useful solutions for the safe buffer overlap function, do not need to be exact, lower bound estimators will not break the operation while still reducing memory use. This means that analytical solutions can simplify certain details for convenience and still be of use.

The memory access behaviour of an operation is distilled into two functions $minR(i)$ and $maxW(i)$ where $i$ is equivalent to $Steps$ as defined in the algorithmic method. These two function $minR(i)$ and $maxW(i)$ have the same meaning as the arrays defined in the algorithmic method, this apporach however will derive equations for them. Figure \ref{fig:depthwise_conv_minr_bound_fn} shows an example of the derived monotonic $minR(i)$ function for depthwise 2D convolution, it can be seen than all read operations, shown in blue, are bounded by the function shown in green.

Using these two functions $O_s$ can then be found using Equation (1) where $i_c$ is the total number of iterations, $OB_s$ is the size of the output buffer, and $T_s$ is the tensor element size in bytes. Strictly this is enough information for a developer to be able to derive an analytic solution for $O_s$ for any algorithm. However for clarity the derivation of the $O_s$ equation for a reference depthwise 2D convolution is shown to illustrate how this can be applied to typical layer implementations used in ML models.

\begin{equation}
O_s = OB_s + min\{minR(i) - maxW(i):i\in\mathbb{Z}\wedge0\geqslant i\geqslant i_c\}T_s
\end{equation}\label{eq:os_analytic}

The approach to solving $O_s$ here is the similar to the algorithmic method, except given analytic solutions for both $minR(i)$ and $maxW(i)$ then Equation (1) can be simplified to an analytic solution. Using these analytic solutions $O_s$ can be computed directly without needing to loop through a large simulated tensor operation, potentially taking millions of iterations.

\begin{figure}[!t]
\centering
\includegraphics[width=2.5in]{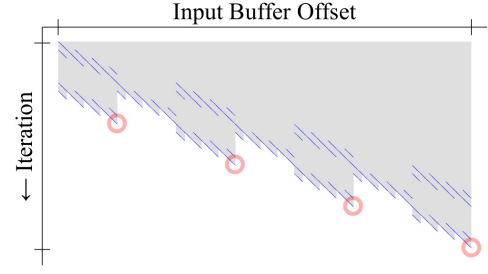}
\caption{Memory read pattern example from a depthwise 2D convolution. Points highlighted define a linear boundary containing all read operations.}
\label{fig:depthwise_conv_reads_highlighted}
\end{figure}

Taking the algorithmic solution for $O_s$ of the reference DepthwiseConv2D operation used above we can study it and derive a purely analytical solution for the lower bound of \begin{math}minR(i)\end{math}. Figure \ref{fig:depthwise_conv_reads_highlighted} shows the pattern of memory reads for an instance of this layer operation, the reads highlighted with red circles can be used to define a linear function encompassing all reads of this algorithm. Firstly the location of the highlighted operations within the loops of the algorithm must be determined. Secondly functions for the iteration and read offset of this points must be produced. 

In the case of this operation the exact read operations highlighted occur during every iteration of the $outY$ loop and the final combined iteration of the $outX$, $ic$ and $m$ loops. Knowing this we determine that each of the highlighted reads occurs where $outY = N, outX = outputW-1, ic = inputD-1, m = filterC-1$. Next we determine that the minimum read within this iteration will always occur when $filterY = 0, filterX = 0$.

Using these observations we can define equations for location of these points in iterations ($i$) and memory offset ($o$) in terms of the iteration through the $outY$ loop ($N$). This is done by tracing these calculations backwards through the original source code to the inputs of the layer.

\begin{equation}
i = (N \cdot O_wO_dK_c) - 1
\end{equation}

\begin{equation}
o = Offset(N\cdot S_w - P_w, (O_w-1)S_h - P_h, I_d - 1)
\end{equation}

Where:

\begin{equation}
Offset(r, c, d) = (r\cdot I_w + c)I_d + d)
\end{equation}
\begin{equation}
P_h = \Bigg \lfloor\frac{O_hS_h - S_h + K_hD_h - D_h - I_h + 1}{2}\Bigg \rfloor
\end{equation}
\begin{equation}
P_w = \Bigg \lfloor\frac{O_wS_w - S_w + K_wD_w - D_w - I_w + 1}{2}\Bigg \rfloor
\end{equation}

Where $I_w$ \& $I_h$ are input shape, $O_w$ \& $O_h$ are output shape, $K_w$, $K_h$ \& $K_c$ are kernel size, $S_w$ \& $S_h$ are stride steps, $D_w$ \& $D_h$ are dilation ratios.

The equations for these points (2) \& (3) can then be used to find the gradient $\frac{di}{do}$ of the line ($a$) and the offset at iteration zero ($b$) defining a linear function which bounds all the read operations of this layer implementation. Simplifying these gives the equations (7) \& (8) for a \& b respectively.

for the gradient and offset of this linear function with reference to the source code of it's implementation:

\begin{equation}
a = \frac{S_h I_w}{O_w K_c}
\end{equation}

\begin{equation}
b = (O_wS_w - P_hI_w - S_h I_w - S_w - P_w + 1)I_d 
\end{equation}

Truncating this linear function at zero gives a good lower bound approximation of the ideal \begin{math}minR(i)\end{math} as shown in Figure \ref{fig:depthwise_conv_minr_bound_fn}.

\begin{equation}
minR(i) = max( 0, a\cdot i + b )
\end{equation}\label{eq:minr}

\begin{figure}[!t]
\centering
\includegraphics[width=2.5in]{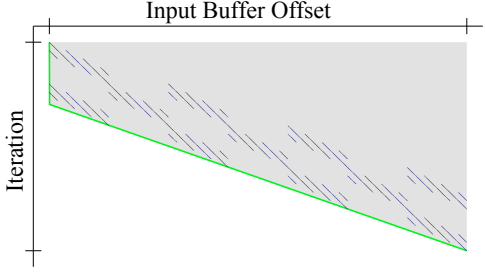}
\caption{$\protect minR(i)$ bounding function for the depthwise 2D convolution implementation. It can be seen that all read operations (in blue) lie above the monotonic function (green).}
\label{fig:depthwise_conv_minr_bound_fn}
\end{figure}

The function \begin{math}maxW(i)\end{math} is trivial in the case of this operation, since each iteration calculates a single element of the output tensor and the loops are nested in increasing dimension order, therefore.

\begin{equation}
maxW(i) = i
\end{equation} \label{eq:maxw}

\begin{figure}[!t]
\centering
\includegraphics[width=2in]{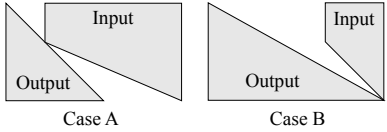}
\caption{The two possible definitions of the analytical minimum bound, depending on the relative gradient of the $minR$ \& $maxW$ functions.}
\label{fig:case_a_b}
\end{figure}

Equations (1), (9) \& (10) can be combined into a single Equation (11) by observing that $O_s$ is defined by the $minR$ \& $maxW$ functions in only two possible ways, as shown in Figure \ref{fig:case_a_b}. If the gradient of $maxW$ is lower than that of $minR$ then it is defined as in case A otherwise it is defined as in case B. These two cases result in the two terms of the $min$ function in the simplified analytical solution for $O_s$ shown in Equation (11). This part of the analytical solution can be used to describe $O_s$ for a wide range of tensor operations, including 2D convolution, all pooling operations as well as the depthwise 3D convolution described here. The only variation of this form are the equations for $a$ and $b$ which define the truncated linear bound of their read offsets.

\begin{equation}
O_s = OB_s + min\bigg\{\frac{b}{a}, ai_c+b-i_c\bigg\}T_s
\end{equation}\label{eq:os_general}

Using this same process the minimum bound linear functions of several common ML reference operations have been derived. Combining these with Equation (10) gives their respective analytical solutions for $O_s$.

Reference 2D Convolution Operation:

\begin{equation}
a = \frac{S_h I_w I_d}{O_w O_d}
\end{equation}

\begin{equation}
b = (O_w S_w - P_h I_w - S_h I_w - S_w - P_w)I_d + 1
\end{equation}

Reference Pooling Operations (all types):

\begin{equation}
a = \frac{S_h I_w}{O_w}
\end{equation}

\begin{equation}
b = (O_w S_w - P_h I_w - S_h I_w - S_w - P_w)I_d + 1
\end{equation}

Example values of $O_s$ calculated using the algorithmic and analytic methods with reference to the ML models analysed are presented in Section \ref{sec:analytic_precision}. It was found that the approximations of the analytic method had a penalty below 2\% of the memory saved and is some cases had no penalty at all. 

\subsection{Precision of the Analyical Method}\label{sec:analytic_precision}
Since the analytical solutions presented in Section \ref{sec:analytical} are lower bounds as opposed to the exact values computed by the algorithmic method in Section \ref{sec:algorithmic}, it is important to quantify the difference between the two. Using MobileNet v2 1.0 224 as an example when diagonal memory optimisation is used, it's peak memory requirement is defined by the second depthwise 2D convolution operation described in Table \ref{tab:conv_details}.

\begin{table}[!t]
\renewcommand{\arraystretch}{1.3}
\caption{Specification of 2nd Depthwise 2D Convolution in MobileNet}
\label{tab:conv_details}
\centering
\begin{tabular}{c||c}
\hline
\bfseries Setting & \bfseries Value\\
\hline\hline
input shape (w, h, c) & 112, 112, 96\\
filter shape (w, h, x, y) & 3, 3, 96, 1\\
output shape (w, h, c) & 56, 56, 96\\
stride (w, h) & 2, 2\\
dilation (w, h) & 1, 1\\
\hline
\end{tabular}
\end{table}

Computing the $O_s$ of this operation using the algorithmic method gives a result of 1204224 bytes, while computing it using the analytic solution presented in Equations (7) (8) \& (11) gives a result of 1193376 bytes. In this case the exact value has been underestimated by 10848 bytes or 0.18\%, in the context of the models optimised memory requirement of 4.6 MB we consider this to be an acceptable approximation.

Table \ref{tab:O_s_estimates} shows these same results for three networks, comparing the exact algorithmic result with the lower bound from the analytical method. The underestimation of $O_s$ ranges from $0\%$ to $0.18\%$. Networks present in the full results but omitted from this table were optimised by reducing element-wise operations, therefore the lower bound approximation had no effect.

\begin{table}[!t]
\renewcommand{\arraystretch}{1.3}
\caption{Estimation Error of Safe Overlap ($O_s$)}
\label{tab:O_s_estimates}
\centering
\begin{tabular}{c||c|c|c}
\hline
\centering\multirow{2}{*}{\bfseries Model} & \multicolumn{2}{c|}{\bfseries Safe Offset ($O_s$)} & \multirow{2}{*}{\bfseries Error}\\\cline{2-3}
& Exact & Estimate\\
\hline\hline
mobilenet v1 1.0 224 		  & 1204224 & 1193376 & 0.18\% \\
mobilenet v2 1.0 224 		  & 1605632 & 1598400 & 0.15\%\\
Inception ResNet v2 		  & 2746884 & 2746884 & 0\%\\
\hline
\end{tabular}
\end{table}

\subsection{Performance Optimised Layer Implementations}\label{sec:effects_of_opt}
The reference depthwise 2D convolution used as an illustrative example in sections \ref{sec:algorithmic} and \ref{sec:analytical} is not the most computationally efficient implementation. This implementation is commonly used for smaller models used in embedded ML applications however more efficient versions are increasingly being used that are optimised for specific processor families, such as cmsis-nn from ARM \cite{lai2018cmsis}. It is important to determine if safe buffer overlapping is possible when using these faster implementations, and to discover if the process described above to derive the analytical solution to $O_s$ is still valid.

Two optimisation approaches are commonly used in embedded ML: loop un-rolling and vectorisation takes advantage of single instruction multiple data (SIMD) operations and reduces loop overheads; while multi-threading can be used when computations as opposed to memory access are the performance bottle-neck. These two approaches are not mutually exclusive and are often used together.

Vectorisation compliments diagonal memory optimisation because multiple elements are processed in longer words, therefore some reads occur earlier and some writes occur later. However it is always possible that trailing elements may need to be computed individually, for this reason the value of $O_s$ for vectorised optimisation is the same as for the reference operations already described.

\begin{figure}[!t]
\centering
\includegraphics[width=2.5in]{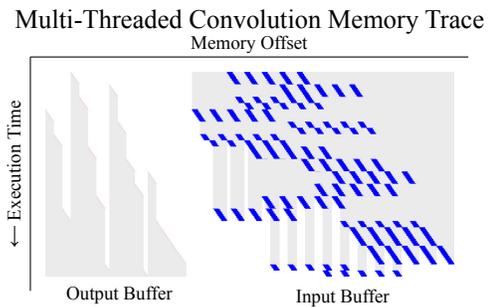}
\caption{Memory trace of a $5 \times 5$ 2D Convolution operation being executed using four threads.}
\label{fig:multithread_trace}
\end{figure}

Multi-threading as it is usually implemented reduces the utility of using diagonal memory optimisation. Threads are generally each given a different contiguous region of the output buffer to compute, resulting in a memory access pattern similar to the one shown in Figure \ref{fig:multithread_trace}. It must be noted that the Valgrind tool used to generate this trace interleaves threads on a single core so does not precisely reproduce true multi-threaded behaviour. However two important features can be seen, firstly four different regions of the output buffer are being computed at similar times and secondly the read/write pattern has become non-deterministic. It is possible to overcome these problems by interleaving the elements each thread computes and ensuring that threads are synchronised to within a maximum offset. It is therefore theoretically possible to write buffer-overlap safe multi-threaded layer implementations, however more work is needed to demonstrate the practicality and reliability of the technique in this case.

\section{Results}\label{sec:results}
A range of network models were analysed with particular emphasis on smaller models intended for mobile applications. These models were sourced from the public repositories Keras Application \cite{keras_apps} and TensorFlow models \cite{tf_models}. Six purely sequential models were analysed, variants of both MobileNet v1 and v2. The sequential nature of these models makes them more amenable to optimisation than the more complex connected networks which have been analysed. Inception v4, Inception ResNet, Nasnet Mobile, DenseNet and ResNet 50 are all complex connected networks, as will be shown this makes predicting the effect of diagonal memory optimisation harder than for simpler sequential models.

A Modified heap allocation algorithm was used to place the tensors buffers in memory for each test and thereby find the peak memory requirement for each model. This algorithm uses conventional heap allocation to place tensor buffers in memory, but chooses a heuristic order in which to allocate buffers which has been found to reduce the peak memory requirement. The next buffer to allocate is chosen using two steps. First the set of un-allocated tensors who’s scope overlap with allocated buffers is found. Out of this set buffer is chosen which can be `heap` allocated into the lowest address space. This algorithm is initiated by allocating a single input or output buffer at offset zero, to perform a forwards or backwards allocation respectively. Each model's tensor graph was serialised using both an eager and lazy execution strategy with the lowest peak memory figure being taken.

The greatest memory savings can be seen in the variations of MobileNet v1 and inception resnet v2 \cite{szegedy2017inception}. In both these models the memory saving occurs in the first few operations where the graph is sequential so the only intermediate values needed are the input and output of a single operation. Again in both cases the optimised operation is a 2D convolution which produces an output tensor twice the size of its input tensor. Here diagonal memory optimisation is able to overlap these two buffers by a few bytes less than size of the input buffer, explaining why the memory saving is almost exactly a third. MobileNet v2 variants also have sequential graphs but in this case the depthwise 2D convolution where the peak memory requirement occurs has an input tensor four times larger than its output. The input and output buffer of this operation are overlapped by almost the full size of the output tensor resulting in a 20\% saving.

\begin{figure}[!t]
\centering
\includegraphics[width=3in]{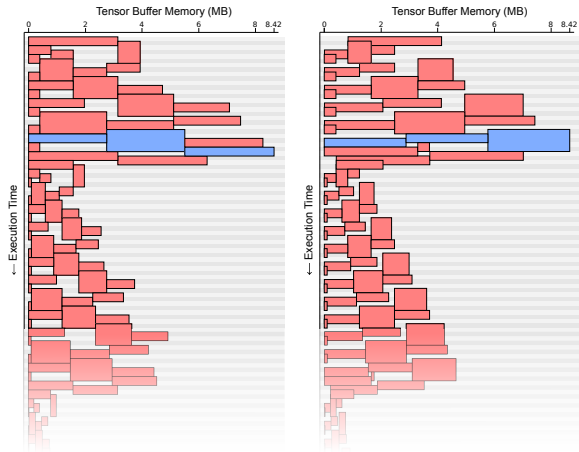}
\caption{a, DenseNet original buffer allocation pattern (only the first fifth of the model is shown for clarity). b, Buffer allocation pattern optimised using DMO, Peak memory defining buffers are shown in blue. It can be seen that none of the buffers which define peak memory are overlapped.}
\label{fig:densenet_pre_post_op}
\end{figure}

The 4.55\% memory saving for DenseNet produced by DMO is an anomaly. In this instance the saving is not produced by diagonal optimisation directly but by a more optimal layout of non-overlapped buffers produced by the heap allocation strategy used. In this case DMO has altered the order that these non-overlapped buffers are allocated so they take up less memory, see Figure \ref{fig:densenet_pre_post_op}. The modified heap allocation strategy used is a heuristic with no guarantee of optimality due to the NP-hard nature of the buffer allocation problem. It is possible that a more effective heap allocation strategy could produce a buffer pre-allocation pattern with this same peak memory requirement without the use of DMO.

The remaining networks Nasnet Mobile and ResNet 50 have the same peak memory requirement even when diagonal memory optimisation is used, both these models are densely connected. Many of the operations within these models produce tensors which are used by more than one subsequent operation, the requirement for these tensor values to be held in memory for longer reduces the opportunities to use DMO which overwrites the values of input buffers.

\begin{table}[!t]
\renewcommand{\arraystretch}{1.3}
\caption{Memory Saving Using Diagonal Optimisation}
\label{tab:results}
\centering
\begin{tabular}{c||c|c|c}
\hline
\centering\multirow{2}{*}{\bfseries Model} & \multicolumn{2}{c|}{\bfseries Peak Memory (KB)} & \multirow{2}{*}{\bfseries Saving}\\\cline{2-3}
& Original & Optimised\\
\hline\hline
MobileNet v1 1.0 224 		  & 4704 & 3136 & 33.3\% \\
MobileNet v1 1.0 224 (8 bit)  & 1176 & 784  & 33.3\%\\
MobileNet v1 0.25 224 		  & 1176 & 786  & 33.2\%\\
MobileNet v1 0.25 128 (8 bit) & 96   & 64   & 33.1\%\\
MobileNet v2 0.35 224 		  & 2940 & 2352 & 20\%\\
MobileNet v2 1.0 224 		  & 5880 & 4704 & 20\%\\
Inception v4 \cite{szegedy2017inception} & 10879 & 10079 & 7.35\%\\
Inception ResNet v2 \cite{szegedy2017inception}		  & 8399 & 5504 & 34.4\%\\
Nasnet Mobile \cite{zoph2018learning} 				  & 4540 & 4540 & None\\
DenseNet 121 \cite{iandola2014densenet} 				  & 8624 & 8232 & 4.55\%\\
ResNet 50 v2 \cite{szegedy2017inception}				  & 10976 & 10976 & None\\
\hline
\end{tabular}
\end{table}

It is important to note that the required memory figures shown in Table \ref{tab:results} only include intermediate tensor values and not the weights of the model itself. In all the models analysed the model weights require significantly more storage than the intermediate values themselves, MobileNet v1 0.25 224 for example has a diagonal memory requirement of 786 KB but has aproximately 2.5 MB of weights. This would make it seem as though diagonal memory optimisation is not be much use in the real world, except that micro-controllers almost universally have much more flash memory than SRAM. The STM32F103xF from ST Microelectronics \cite{stm32} is a commonly used ARM Cortex M3 micro-controller with 768 KB or 1 MB of program storage and 96 KB of SRAM. Using diagonal memory optimisation it becomes possible to execute the smallest MobileNet (v1 0.25 128 8bit) on this chip, however the weights of this model take 623 KB, 60.8\% of the micro-controllers program memory. Similarly the Atmel AT32UC3C \cite{at32} used by the on-board computer of the ESAs ESEO mission \cite{bartram2017software} has at least four times more flash memory than SRAM across all its variants.

\section{Conclusion}\label{sec:conclusion}
An opportunity for memory optimisation was discovered while inspecting the memory use patterns of compiled ML models, this opportunity to reduce the RAM requirements of running these models is especially important in edge ML applications where models are running on low power micro-controllers. 
The method of diagonal memory optimisation has been described and several methods presented to compute the critical buffer overlap $O_s$ metric required to use this technique. Including a formal analytic approach to easily compute the lower bound of $O_s$. Several other techniques with the potential to reduce the memory requirement of ML models have been discussed, and it has been shown that they are complimentary with diagonal memory optimisation itself.

The operation splitting technique described here has been shown to be possible, but further work is needed in this area to formalise its use. Additional work is needed in order to use diagonal memory optimisation with multi-threaded implementations of ML models where thread synchronisation is critical to avoid computed values in memory being overwritten.

Significant memory savings have been demonstrated on a range of real world models including making the smallest possible implementation of MobileNet even smaller. The memory optimisation algorithm described in this paper is available in the open-source TFMin tool \cite{tf_min}, to ease its investigation and adoption by the community. Diagonal memory optimisation will enable larger ML models to be executed on micro-controller targets than currently possible, enabling smarter devices and or savings in power and cost.

\appendices



\ifCLASSOPTIONcaptionsoff
  \newpage
\fi



%
\bibliographystyle{IEEEtran}
\bibliography{IEEEabrv,References}

\end{document}